\documentclass[fleqn,12pt,twoside]{article}
\usepackage{espcrc1}

\usepackage{graphicx}
\usepackage[figuresright]{rotating}

\newcommand{\AmS}{{\protect\the\textfont2
  A\kern-.1667em\lower.5ex\hbox{M}\kern-.125emS}}

\title{Parity-Dependence in the Nuclear Level Density}

\author{D. Mocelj\address[BS]{Department for Physics and Astronomy, University of Basel,
        Basel, Switzerland}\thanks{This work was supported by the Swiss NSF grants 2000-061031.02 and 2024-067428.01}
      , T. Rauscher\addressmark[BS], K. Langanke\address[Aar]{Institute for Physics
        and Astronomy, University of Aarhus, Aarhus, Denmark}, G. Mart{\'i}nez Pinedo\address[BCN]
      {Instituci{\'o} Catalana de Recera i Estudis Avan{\c{c}}ats, Barcelona, Spain}, 
      L. Pacearescu\address[Tueb]{Institute of Theoretical Physics, University of T\"ubingen,
        T\"ubingen, Germany}, \\
      A. F\"a{\ss}ler\addressmark[Tueb], and F.-K. Thielemann\addressmark[BS]}
       
\begin{document}

% typeset front matter
\maketitle

\begin{abstract}
Astrophysical reaction rates are sensitive to the parity distribution at low excitation energies. 
We combine a formula for the energy-dependent parity distribution with a microscopic-macroscopic nuclear level 
density. This approach describes well the transition from low excitation energies, where a single parity dominates, to 
high excitations where the two densities are equal.
\end{abstract}

\section{Introduction}
The nuclear level density is an important ingredient in the prediction of nuclear reaction rates in astrophysics. 
So far, most theoretical, global calculations of astrophysical rates assume an equal distribution of the state 
parities at all energies. It is obvious that this assumption is not valid at low excitation energies of a nucleus. 
However, a globally applicable recipe was  lacking. For nuclei far from stability, where no experimental information 
on excited states is available, a large effect of the parity dependence on 
predicted cross sections can be expected. 
 
\section{Method}

\begin{figure}[t]
  \includegraphics[keepaspectratio,height=10cm,angle=-90]{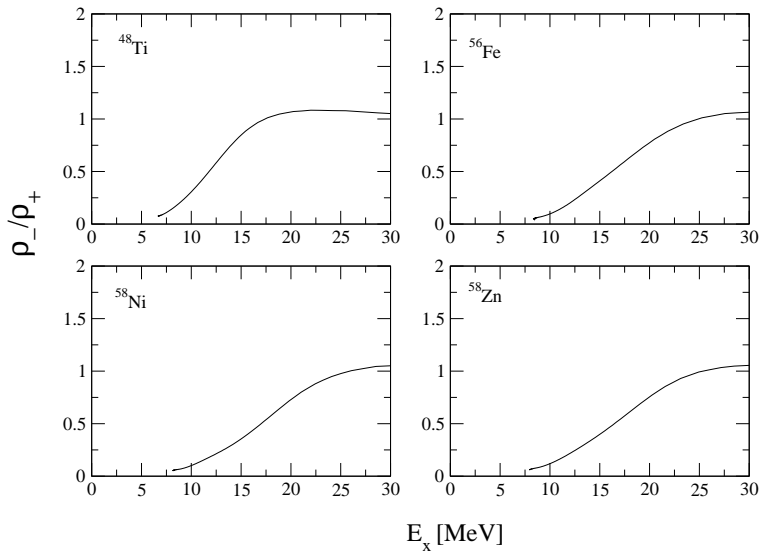}
  \caption{Odd- to even parity ratio calculated for $^{48}$Ti, $^{56}$Fe, $^{58}$Ni and $^{58}$Zn}
  \label{fig:1}
\end{figure}

\begin{figure}[t]
  \includegraphics[width=0.5\textwidth,angle=-90]{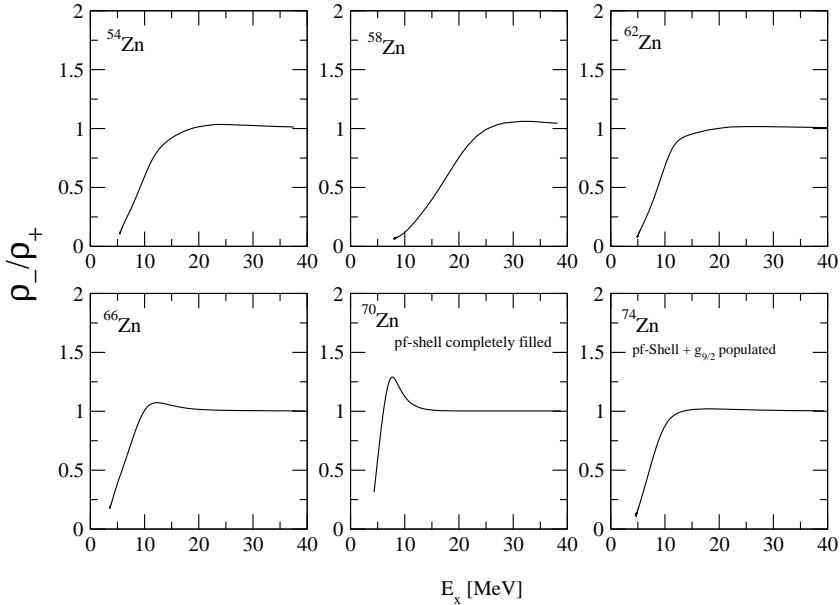}
  \caption{Evolution of the odd- to even-parity ratio within the Zn-chain.}
  \label{fig:2}
\end{figure}

\begin{figure}[t]
  \includegraphics[width=0.5\textwidth,angle=-90]{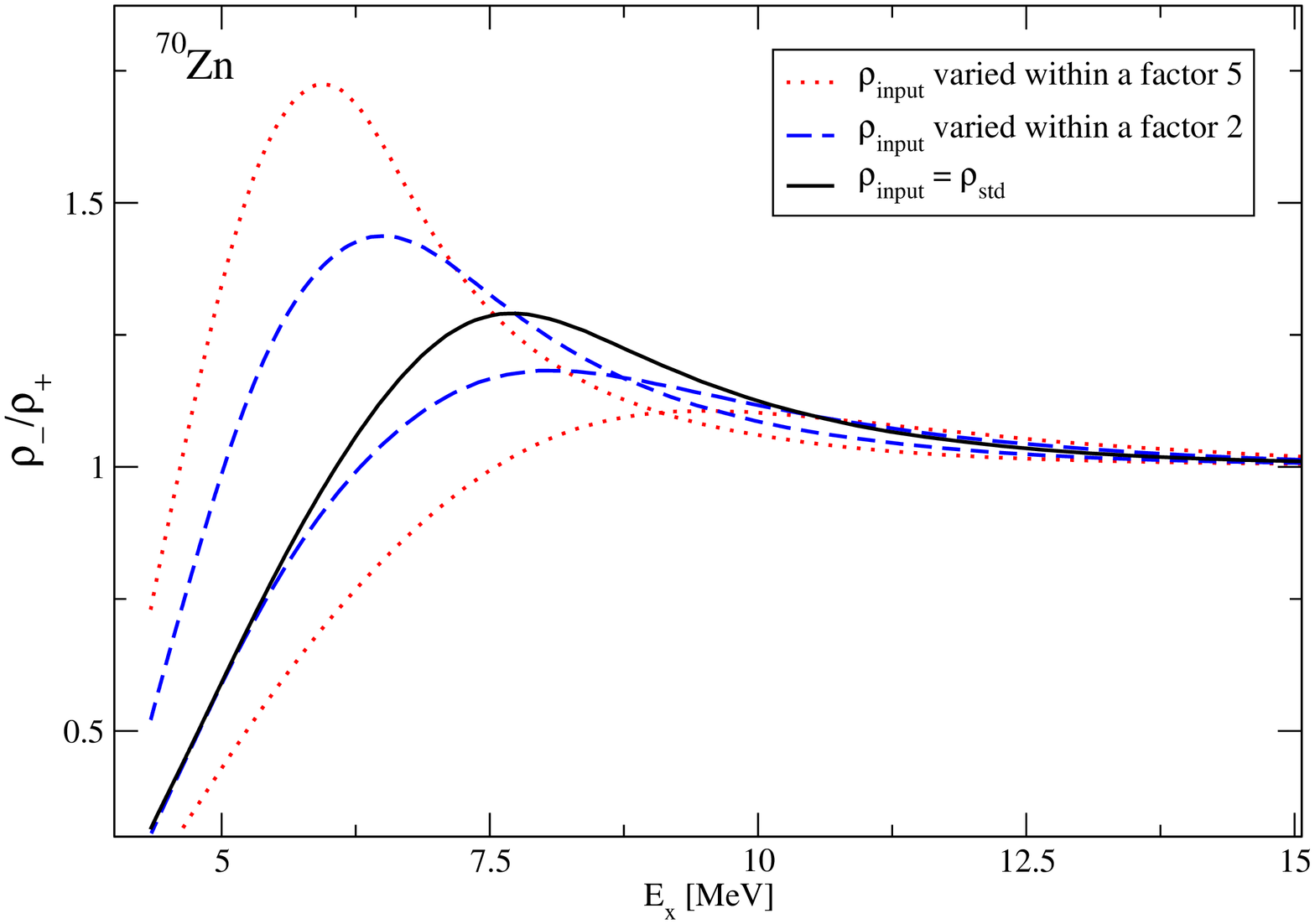}
  \caption{Influence of the input level-density parameter $a$ on the peak strength in $^{70}$Zn. The black curve is
  obtained by using the standard level-density parameter \cite{tommy}. The dashed (dotted) curve corresponds 
  to a variation of the level-density parameter $a$ which translates to a variation of the input level density by a 
  factor of 2 or 5, respectively.}
  \label{fig:3} 
\end{figure}

Single particle levels are divided into two groups, according to the individual parities. The group which has the 
smaller average occupation number is denoted by $\Pi$. Assuming that nucleons occupy the single-particle orbitals 
independently and randomly, the occupancy $n$ of the $\Pi$-parity orbits is given \cite{alha} by a Poisson 
distribution,
\begin{equation} 
P(n)= \frac{f^n}{n!}e^{-f}
\end{equation} \label{eq:1}
where $f$  is the average occupancy of orbits  with parity $\Pi$. Then the ratio of the odd-parity to the even-parity 
probabilities is given by \cite{alha}
\begin{equation}
\frac{P_-}{P_+}=\frac{Z_-}{Z_+}= \tanh f. 
\end{equation}
In nuclei, $f$ has to be replaced by the sum of individual contributions from neutrons and protons. 
To calculate the total partition function $Z$ we use the macroscopic-microscopic nuclear level density of \cite{tommy}.
The average occupancy $f$ is computed from BCS occupation numbers based on single particle levels from an axially 
symmetric deformed Saxon-Woods potential \cite{wscode} with parameters from \cite{wspara} which reproduce experimental 
data well \cite{exp1,exp2}. All major shells up to $11\hbar\omega$ were included which allows to extend our 
calculations way beyond the previously studied $pf+g_{9/2}$ shell.
Using $Z_++Z_-=Z$ and Eq.\ \ref{eq:1}, we can thus determine $Z_\pm$ and calculate the thermal energies for even- and
 odd parity states. Canonical entropies, heat capacities, and the parity projected level densities at a given 
excitation energy, are derived from standard thermodynamic relations. 

\section{Results and Discussion}

Typical results for nuclei in the Fe-region are shown in Fig.\ 1. 
On can see that the assumption of equally distributed state parities is not fulfilled. Even at excitation energies of 
10 - 15 MeV, the parity ratio is not yet equilibrated. 

The evolution of the parity ratios within an isotopic chain is shown in Fig. 2.
Starting with $^{54}$Zn, where the $pf$-shell is filled in neutrons only to 20 \%, and stopping with $^{74}$Zn, 
where the next major shell has started to be populated, one can  see that the ratio approaches unity for lower values
 of the excitation energy as one approaches the $N=40$ shell closure.  As the parity can only be changed by 
excitations from the $pf$  to the $sdg$ shell, the ratio will equilibrate faster with increasing neutron number 
as the gap between the last occupied orbit in the $pf$-shell and the $sdg$-shell will decrease. For $^{70}$Zn, where 
the $pf$-shell is completely filled, a pronounced peak around 6 MeV shows up, which might be understood as follows: 
As the $pf$-shell is completely filled, the parity of the system will be changed by any neutron excitation, 
resulting in a dominance of negative parity states at the energies for which the peak appears. However, as shown 
below, this effect strongly depends on the inputs and has to be interpreted with caution.

There are two essential inputs to our calculations: the total level density $\rho_\mathrm{tot}$ which is used to 
calculate
the total partition function utilized in Eq.(\ref{eq:1}) and the single particle levels from the deformed Saxon-Woods
potential needed to compute the average occupancy $f$. Both inputs are prone to possible uncertainties. We explored
the sensitivity to these uncertainties by simulating the combined effects of uncertainties in both inputs by variation
of only the total level density $\rho_\mathrm{tot}$ and keeping the Saxon-Woods parameters unchanged. Fig.\ 
\ref{fig:3} shows the impact of varying the level density parameter $a$ in the range of $8\leq a\leq 12$ MeV$^{-1}$.
This corresponds to a $\pm 20$ \% variation of the standard parameter $a$ and implies
a change in $\rho_\mathrm{tot}$ of up to a factor of 5 at an excitation energy
of 6 MeV. It has to be emphasized
once more that this large variation of the level density is not due to the uncertainty of $\rho_\mathrm{tot}$ alone 
but rather is supposed to contain the combined uncertainties in $\rho_\mathrm{tot}$ {\em and} the single particle 
levels. Fig.\ \ref{fig:3} clearly shows the need for improved consistency within the inputs, especially at shell 
closures.

\section{Conclusion and Outlook}

We have shown that the assumption of equally distributed state parities is only justified for high excitation 
energies. For lower energies, even still at particle separation energies which are in the order of several MeV, this 
assumption is clearly not fulfilled. 
Further investigations are needed to arrive at an understanding of the peak-strengths found at shell closures. Work 
is in progress to calculate the parity distribution for a large number of nuclei far from stability on the proton 
as well as neutron rich side. Influences on nucleosynthesis  calculations will be studied.

\end{document}